\documentclass[a4paper]{article}
\usepackage[T1]{fontenc}
\makeatletter

\newcommand{\LyX}{L\kern-.1667em\lower.25em\hbox{Y}\kern-.125emX\spacefactor1000}

\newcommand{\lyxaddress}[1]{
  \par {\raggedright #1 
  \vspace{1.4em}
  \noindent\par}
}

\makeatother

\begin{document}

\title{An Exact Determination of the Electron Exchange-Correlation of \protect\( ^{9}Be\protect \),
\protect\( ^{10}B\protect \), \protect\( ^{12}C\protect \)}

\author{J.A. de Wet}

\maketitle

\lyxaddress{\centering Box 514, Plettenberg Bay, 6600, South Africa}

\begin{abstract}
We show in this note how many electron irreducible representations of the Lorentz
group L can be expressed in terms of Slater determinants. In this way the full
configuration wave function of Quantum Chemistry is derived without any appeal
to the Schrodinger equation and Born-Oppenheimer approximation. In particular
the number \( n \)of exchange-correlations between the orbiting electrons and
those in the neutrons may be calculated and therefore the ground state electron
exchange-correlation energy \( E \) readily obtained by dividing the atomic
energies by \( n \). Comparison with measured atomic energies yields values
of \( E \) very close to -1 hartrees for the 3 neutral atoms \( Be \), \( B \)
and \( C \). This tallies within a few percent with the best available \( E \)
estimates. Although calculations were performed for 3 atoms the method is perfectly
general and can be applied to any atom. However because \( E \) is estimated
for metals only, gases are not considered in this paper. The intimate relationship
with density function theory is discussed in the conclusion.
\end{abstract}

\section{Introduction}

Many years ago the many electron problem was modelled by irreducible representations
of the Pauli ring found by constructing the minimal left ideals \cite{1}. Then
when a field was introduced by fibration the well known spin angular momentum
matrices appeared. The minimal left ideals are the bra vectors of a density
operator \( \psi  \) \cite{2}, so the method is in fact a density function
formulation as well as a representation of the two-dimensional unitary group
\( U_{2} \) and therefore of a subgroup of the proper orthochronous homogeneous
Lorentz group \( L_{p} \). In this paper the many electron case will be extended
to the nucleus by considering the neutron to be a proton plus an electron so
a neutral atom is envisaged as an ensemble of A protons and A electrons distributed
between N neutrons and A-N orbiting electrons. Under these conditions the complex
angular momentum matrix is \cite{3}.

{\centering \( \gamma =E_{N}\otimes \gamma _{P}+\gamma _{N}\otimes E_{P} \)\marginpar{
(1)
}\par}

Where \( \gamma _{P} \), \( \gamma _{N} \) are \( (P+1)- \)and \( (N+1)- \)dimensional
Lie operators of SO (3) and \( E_{P} \), \( E_{N} \) are \( (P+1) \) and
\( (N+1) \) unit matrices. However the nett electron spins must also be matched
with the spin of the nuclear ground state even though the spins may be orthogonal
and therefore not additive.

According to this scenario the exchange energy of interaction responsible for
atomic energies will be between the \( P \) outer electrons and the \( P \)
electrons within the nucleus and thus the number of interactions will be of
the order of \( P^{2} \) unless they occur simultaneously. It will be seen
that this is indeed supported by experimental evidence. Furhermore a preferred
spin orbital \( \psi _{e} \) will be found for the collective motions of the
orbiting electrons.

In order to actually calculate the number of exchange-correlations associated
with \( \psi _{e} \) it is necessary to integrate, by exponentiation, the representation
of the Lie algebra \( u_{2} \) characterising the electronic spin state matching
the nuclear ground state. When this is done, by means of a matrix exponential
theorem, the Slater determinants incorporating exchange correlations emerge
naturally.

To illustrate the method, which is quite general, we will begin with \( ^{9}Be \)
and then in Section 3 go on to \( ^{10}B \) and \( ^{12}C \) which belong
to another space because \( A \) is even. Specifically the bosons (even \( A \))
are characterised by Euclidean structure while the fermions belong to complex
manifolds.

\section{BERYLLIUM-9}

According to de Wet \cite{1} the many electron configuration decomposes into
the sum \( \sum C_{\lambda }P_{\lambda } \) of projection operators \( P_{\lambda } \)
characterising the spin state \( \left[ A-\lambda ,\lambda \right]  \) and
coefficients \( C_{\lambda } \) which are polynomials in \( \gamma  \) of
equation (1) and may be determined by the recurrence formula:

\( \left( n+1\right) C_{n+1}=\gamma C_{n}+\left( A-(n-1)\right) C_{n-1} \),
\( n\geq 1 \), \( A>(n-1) \)\marginpar{
(2)
}

where

\( \gamma =C_{1}=E_{n}\otimes \gamma _{P}+\gamma _{N}\otimes E_{P} \), \( C_{0}=1 \)\marginpar{
(1)
}

is related to the usual spin matrices by \( \gamma =2is \). Thus we find

\( n=1 \), \( C_{2}=\frac{1}{2}\left( \gamma ^{2}+A\right)  \)\marginpar{
(2a)
}

\( n=2 \), \( 6C_{3}=\gamma \left( \gamma ^{2}+(3A-2)\right) =\gamma ^{3}+25\gamma  \)

\( =E_{5}\otimes (\gamma ^{3}_{P}+25\gamma _{P})+3\gamma ^{2}_{N}\otimes \gamma _{P}+3\gamma _{N}\otimes \gamma ^{2}_{P}+(\gamma ^{3}_{N}+25\gamma _{N})\otimes E_{4} \)\marginpar{
(2b)
}

when A=9. Here \( C_{3} \) defines the spin state \( \frac{3}{2} \)corresponding
to 6 electrons with positive spin and 3 with negative spin, thus matching the
spin of the ground state of \( ^{9}Be \) nucleus where P=4, N=5. In this way
the electronic and nucleon spins are linked even though their spins may be orthogonal
and therefore not additive.

If we use the diagonal representations

\( i\gamma _{P}=\left\{ 4;2;0;-2;-4\right\}  \), \( i\gamma _{N}=\left\{ 5;3;1;-1;-3;-5\right\}  \)\marginpar{
(3)
}

then \( C_{3} \) can easily be obtained from (2b). Half the values appear in
Table I; the remainder are repeated with a change of sign.

\textbf{TABLE I: }

\vspace{0.3cm}
{\centering \begin{tabular}{|c|ccccc|ccccc|ccccc|}
\hline 
\( E_{5}\otimes \gamma _{P} \)&
-4i&
-2i&
0&
2i&
4i&
-4i&
-2i&
0&
2i&
4i&
-4i&
-2i&
0&
2i&
4i\\
\( \gamma _{N}\otimes E_{4} \)&
-5i&
-5i&
-5i&
-5i&
-5i&
-3i&
-3i&
-3i&
-3i&
-3i&
-i&
-i&
-i&
-i&
-i\\
\( \gamma  \)&
-9i&
-7i&
-5i&
-3i&
-i&
-7i&
-5i&
-3i&
-i&
i&
-5i&
-3i&
-i&
i&
3i\\
\( \frac{1}{8}C_{3} \)&
\( \frac{21}{2}i \)&
\( \frac{7}{2}i \)&
0&
-i&
\( -\frac{i}{2} \)&
\( \frac{7}{2}i \)&
0&
-i&
\( -\frac{i}{2} \)&
\( \frac{i}{2} \)&
0&
-i&
\( -\frac{i}{2} \)&
\( \frac{i}{2} \)&
i\\
\hline 
\end{tabular}\par}
\vspace{0.3cm}

If however we consider electrons spinning about the \( x_{2} \)- axis then
\[
\left\langle jm\prime \mid i\gamma \mid jm\right\rangle =\left[ \left( j-m\right) \left( j+m+1\right) \right] ^{\frac{1}{2}}\delta _{m\prime ,m+1}\]
\marginpar{
(4)
}where \( i\gamma _{P} \) is characterised by \( j=2 \), \( i\gamma _{N} \)
by \( j=\frac{5}{2} \)and
\[
\frac{1}{8}C_{3}=\left[ \begin{array}{cc}
0 & X\\
-X^{T} & 0
\end{array}\right] \]
\marginpar{
(5)
}

Which belongs to a complex space even though the eigenvalues of X are real.
After symmetrization, by interchanging rows and columns, X is found to have
the eigenvalues of Table I which label its 15 rows, and we will assume from
symmetry considerations that X\( _{77} \) corresponds to the eigenvalues \( \left| \frac{i}{2}\right|  \)which
is the same as the lowest non-zero collective eigenvalue due to 4 spinning protons
and 5 spinning neutrons obtained by the method outlined in \( \left[ 4\right]  \).
This corresponds to identifying the 3 lowest non-zero eigenvalues of \( \frac{C_{3}}{8} \)with
electrons in the nucleus and the high eigenvalue \( 10.5i \) with the collective
motion of the 4 orbital electrons. Then when (5) is exponentiated the Slater
determinants and principal minors governing the number of exchange correlations
of the 4 electrons with those in the nucleus will emerge without any further
assumptions.

The matrix exponential theorem derived by de Wet {[}3, 4{]} says that

\( e^{\mu \theta }=\mu \begin{array}{c}
n\\
\sum \\
k=0,1,..
\end{array}\frac{F_{k}(\mu )}{i\lambda _{k}F_{k}(i\lambda _{k})}\cos \lambda _{k}\theta +i\begin{array}{c}
n\\
\sum \\
k=1,...
\end{array}\frac{F_{k}(\mu )}{F_{k}(i\lambda _{k})}\sin \lambda _{k}\theta  \)\marginpar{
(6)
}

where \( \lambda _{k} \) is the k-th eigenvalue of \( \mu =X \), and

\( F_{k}(\mu )=F(\mu )/(\mu ^{2}+\lambda _{k}^{2}) \), \( F_{j}(\mu )F_{k}(\mu )=0 \)\marginpar{
(6a)
}

automatically selects an occupied orbital representing an excited state {[}k{]}
because

\( F(\mu )=\mu (\mu ^{2}+1)(\mu ^{2}+\lambda ^{2}_{2})\cdots \cdots (\mu ^{2}+\lambda ^{2}_{n})=0 \).\marginpar{
(6b)
}

To see it's physical significance we write

\( F_{k}(\mu )=\mu ^{2n-1}+\beta _{1}\mu ^{2n-3}+\beta _{2}\mu ^{2n-5}+\cdots \cdots \cdots +\beta _{n-1}\mu  \)\marginpar{
(7)
}

where \( \mu ^{3}\rightarrow -\mu ^{3} \), \( \mu ^{5}\rightarrow +\mu ^{5} \),
etc. because (5) belongs to a complex space and

\( \beta _{1}=\lambda _{1}^{2}+\lambda ^{2}_{2}+\cdots \cdots \cdots +\widehat{\lambda _{k}}^{2}+\cdots \cdots \cdots +\lambda ^{2}_{n} \),
\( (\lambda _{k}=0) \)\marginpar{
(7a)
}

\( \beta _{2}=\lambda ^{2}_{1}\lambda _{2}^{2}+\lambda _{1}^{2}\lambda ^{2}_{3}+\cdots \cdots \cdots +\lambda ^{2}_{1}\lambda ^{2}_{n}+\cdots \cdots \cdots +\lambda _{n-1}^{2}\lambda _{n}^{2} \),
\( (\lambda _{k}=0) \)\marginpar{
(7b)
}

\( \vdots  \)

\( \beta _{n-1}=\lambda ^{2}_{1}\lambda ^{2}_{2}\cdots \cdots \cdots \widehat{\lambda _{k}}^{2}\cdots \cdots \cdots \lambda ^{2}_{n}=\det X_{k} \),
\( (\lambda _{k}=1) \)\marginpar{
(7\( _{n} \))
}

Equation (7\( _{n} \)) is a Slater determinant of the possible orbits, excluding
the state {[}k{]}, and the remaining equations are principal minors representing
clusters of correlated electrons \cite{5}. The expansion coefficient in the
full configuration wave function (6) are the matrix powers \( \left[ \mu ^{r}\right] _{ik} \)
which depend on the state {[}k{]} and the spins and rotations associated with
k.

To prove (6), differentiate at \( \theta =0 \) to get

\( \left. \frac{de^{\mu \theta }}{d\theta }\right| _{\theta =0}=\mu =\mu \begin{array}{c}
n\\
\sum \\
k=1,2,..
\end{array}\frac{i\lambda _{k}F_{k}(\mu )}{\mu F_{k}(i\lambda k)} \)\marginpar{
(8)
}

However \( K_{k}(\mu )=\frac{i\lambda _{k}F_{k}(\mu )}{\mu F_{k}(i\lambda k)} \)\marginpar{
(8a)
}

is idempotent so \( \sum K_{k}(\mu ) \) is a decompositon of unity and (6)
follows. The connection between idemponency and statistics will be discussed
in the conclusion.

Returning to (6) the sin term is characterised by odd powers of \( \mu  \)
and belongs to the matrix (5). The cosine term will belong to a Euclidean space
defined by matrices along the diagonal.

In the case of the problem under consideration the independent eigenvalues are

\( \lambda ^{2}=\left\{ \frac{1}{4};1;\frac{49}{4};\frac{441}{4}\right\}  \)

and the relevant matrix powers are

\( X_{77}=\frac{9}{4} \), \( \left[ X^{3}\right] _{77}=269.390625 \), \( \left[ X^{5}\right] _{77}=29888.34375 \),
\( \left[ X^{7}\right] _{77}=3297590.147461 \)

so (6) yields the electronic wave function

\( \psi _{77}=\frac{1}{64}(-22\sin \frac{\theta }{2}+8\sin \theta -3\sin \frac{7}{2}\theta +15\sin \frac{21}{2}\theta ) \)\marginpar{
(9)
}

which satisfies the criterion \( \left. \frac{d\psi _{77}}{d\theta }\right| _{\theta =0}=\frac{9}{4} \).

Comparing (9) with the full configuration interaction wave function proposed
by Szabo and Ostlund ({[}5{]}, Ch. 2) the coefficients of the sine terms are
integral exchange correlations up to a normalization constant (which at least
ensures that one eigenvalue of the set assumes the value of unity as required
by the theorem).

In particular the spin orbital \( \sin (\frac{21}{2}\theta ) \) associated
with the 4 orbiting electrons would have 15 exchange correlations which would
be approximately the case if all exchanges between these electrons and 4 in
the nucleus occured independently implying 16 exchanges. Johnson et al.\cite{6}
list -14.6674 hartrees for the measured exchange-correlation energy of Be which
yields an exchange-correlation energy per electron of -14.6674/15=-0.9778 hartrees.
In the next section (Table II) this value will be compared with the best available
estimate given by the formula of Gell-Mann and Brueckner \cite{7}.

\section{Boron-10, Carbon-12}

The same method may be used with the difference that because we are now dealing
with bosons that occupy Euclidean space, equation (6) assumes the simpler form
\cite{3}.

\( e^{i\mu \theta }=\begin{array}{c}
n\\
\sum \\
k=0,1,..
\end{array}\frac{F_{k}(\mu )}{F_{k}(\lambda k)}e^{i\lambda _{k}\theta } \),\marginpar{
(10)
}

\( F_{k}(\mu )=\mu (\mu -1)(\mu -\lambda _{2})\cdots \cdots (\mu -\lambda _{n})/(\mu -\lambda _{k}) \),
\( F_{j}(\mu )F_{k}(\mu )=0 \)

and

\( \lambda _{k}F_{k}(\mu )/\mu F_{k}(\lambda _{k}) \)\marginpar{
(10a)
}

is idempotent, so that

\( \left. \frac{de^{i\mu \theta }}{d\theta }\right| _{\theta =0}=i\mu \sum \frac{\lambda _{k}F_{k}(\mu )}{\mu F_{k}(\lambda _{k})}=i\mu  \)

In the case of boron we shall employ the coefficient 

\( 24C_{4}=\gamma ^{4}+52\gamma ^{2}+240 \)

\( =E_{5}\otimes (\gamma ^{4}_{P}+52\gamma ^{2}_{P})+4\gamma _{N}\otimes (\gamma ^{3}_{P}+13\gamma _{P})+6\gamma ^{2}_{N}\otimes \gamma ^{2}_{P}+4(\gamma ^{3}_{N}+13\gamma _{N})\otimes \gamma _{P}+(\gamma ^{4}_{N}+52\gamma ^{2}_{N})\otimes E_{5}+240 \)\marginpar{
(11)
}

corresponding to a spin state 1, with the degenerate eigenvalue

\( \lambda =\{210;42;10;2;-14\} \)\marginpar{
(12)
}

which in terms of the theorem have to be normalised so that 1 eigenvalue is
unity and all eigenvalues are positive. To do this we add \( \lambda _{t}=14 \)
and divide by \( \lambda _{f}=16 \) to find the positive canonical form

\( \mu =(C_{4}+14)/16=\{14;3.5;1.5;1;0\} \)\marginpar{
(12a)
}

The constant \( \lambda _{t} \) will translate the spectrum because if \( AX=\lambda X \),
then

\( (A-\lambda _{t})X=(\lambda -\lambda _{t})X \)

And after exponentiation will multiply the wave function by a factor \( e^{i\lambda _{t}\theta } \).
On the other hand \( \lambda _{f} \) may be absorbed in \( \theta  \) and
simply lead to a irrelevant change in frequency that will not affect the coefficients
\( F_{k}(\mu )/F_{k}(\lambda _{k}) \) of the wave function (10) which are independant
of which eigenvalue is normalised. Also the zero in (12a) contributes nothing
to the wave function and the theorem counts degenerate eigenvalues only once.

The \( 36\times 36 \) matrix \( \mu  \) decomposes into two bisymmetric diagonal
\( 18\times 18 \) matrices of which only one gives the required eigenvalues
(12a). By symmetry we choose \( \mu _{99}=\frac{41}{16} \) as the matrix element
corresponding to the ground state and then using (10) find a unitary wave function

\( 256\psi _{99}=\exp (\frac{14}{16}\theta )\{74\sin \theta +38\sin \frac{3}{2}\theta +50\sin \frac{7}{2}\theta +25\sin 14\theta \} \),\marginpar{
(13)
}

\( \left. d\psi _{99}/d\theta \right| _{\theta =0}=\frac{41}{16} \)

Again there is a high frequency spin orbital \( \sin 14\theta  \) associated
with the 5 orbiting electrons with 25 exchange correlations yielding an exchange
correlation energy per electron of \( -24.6539/25=-0.9862 \) hartrees using
the results of \cite{6}.

Finally the coherent electron coefficient corresponding to a zero spin state
of C is

\( 720C_{6}=\gamma ^{6}+140\gamma ^{4}+4144\lambda ^{2}+14400 \)\marginpar{
(14)
}

and \( \mu =1-C_{6}/84 \), has the degenerate eigenvalues

\( \{12;\frac{4}{3};1;\frac{16}{21};0\} \)

belonging to a \( 24\times 24 \) bisymmetric matrix with \( \mu _{12,12}=211/84 \)
labelling the ground state in which (10) yields a wave function

\( 256\psi _{12,12}=\sin \theta \{39\sin \frac{16}{21}\theta +128\sin \theta +26.5\sin \frac{4}{3}\theta +37.5\sin 12\theta \} \)\marginpar{
(15)
}

In this example which is not a metal, it would seem that there is an electron
exchange shared by the adjacent state \( \sin \frac{4}{3}\theta  \) such that
the total number of exhange correlations associated with the outer orbitals
remains constant at \( 26.5+37.5=64 \). If we assume that the largest negative
exchange correlation energy is favoured then the exchange correlation energy
per electron will be \( -37.8450/38=-0.9959 \) hartrees.

We will conclude this section with a comparison of the exact ground state correlation
exchange energies with those derived by Gell-Mann and Brueckner for metals \cite{7}
who give the formula

\( \epsilon _{g}=-\frac{1}{2}\left( \frac{0.916}{r_{s}}-0.0622\ln r_{s}+0.094\right)  \)
hartrees\marginpar{
(16)
}

Where \( r_{s} \) is the inter-electron spacing in units of Bohr radius. Thus
if we take volume occupied by one electron to be \( r_{s}^{3} \) then the inverse
density is \( N^{-1}=r^{3}_{s} \) in these units, where \( N=A \) is the number
of electrons, and (16) becomes

\( \epsilon _{g}=-\frac{1}{2}\left( 0.916N^{1/3}+0.02073\ln N+0.084\right)  \)
hartrees\marginpar{
(16a)
}

Here the constant term has been slightly modified so that \( \epsilon _{g}=-\frac{1}{2} \)
hartrees when \( N=1 \) which is the known value for hydrogen.

Table II compares \( \epsilon _{g} \) values derived from (16a) with those
calculated in sections 2, 3.

Table II

\vspace{0.3cm}
{\centering \begin{tabular}{|c|c|c|}
\hline 
Atom&
Exact \( \epsilon _{g} \)&
\( \epsilon _{g} \) by (16a)\\
\hline 
\hline 
H&
-0.5&
-0.5\\
\hline 
Be&
-0.9778&
-0.9746\\
\hline 
B&
-0.9862&
-1.0476\\
\hline 
C&
-0.9959&
-1.1113\\
\hline 
\end{tabular}\par}
\vspace{0.3cm}

\section{Conclusion}

According to Biedenharn and Louck \cite{2}, Section 7.7 this work is, in fact,
a density functional formalism because the wave function \( \sum C_{\lambda }P_{\lambda } \)
is generated by a left ideal and (10) has the form of a general density matrix
because if \( \lambda _{k}F_{k}(\mu )/\mu F_{k}(\lambda _{k}) \) is idempotent
then so is \( F_{k}(\mu )/F_{k}(\lambda _{k}) \) since \( F_{k}(\mu ) \) vanishes
unless \( \mu =k \). The use of density matrices to describe canonical ensembles
has been detailed by Prigogine \cite{8}. In fact the idempotent matrices are
Slater determinants and principal minors which describe pure or coherent states,
so one is looking at nett many particle behaviour instead of attempting to build
this up from the interaction of individual particles. Specifically we have the
partition

\( (A-\lambda )+\lambda =A \)

into the two states of a canonical ensemble namely; \( (A-\lambda ) \) electrons
with positive spin and \( \lambda  \) electrons with negative spin as it is
possible that very light elements such as Li may only be approximated. The intimate
connection between statistics and idempotency was actually recognised long ago
by Eddington \cite{9}, Section 66.

\end{document}